# Evolution of the magnetic anisotropy with carrier density in hydrogenated (Ga,Mn)As

L. Thevenard*, L. Largeau, O. Mauguin, and A. Lemaître

Laboratoire de Photonique et Nanostructures, CNRS,

Route de Nozay, 91460 Marcoussis, France

K. Khazen and H. J. von Bardeleben

Institut des Nanosciences de Paris, Universités Paris 6&7 / UMR 7588,CNRS

Campus Boucicaut, 140,rue de Lourmel, 75015 Paris, France

The magnetic properties of (Ga,Mn)As thin films depend on both the Mn doping level and the carrier concentration. Using a post growth hydrogenation process we show that it is possible to decrease the hole density from $1.10^{21}$ cm$^{-3}$ to $<10^{17}$ cm$^{-3}$ while maintaining the manganese concentration constant. For such a series of films we have investigated the variation of the magnetization, the easy and hard axes of magnetization, the critical temperatures, the coercive fields and the magnetocrystalline anisotropy constants as a function of temperature using magnetometry, ferromagnetic resonance and magneto-transport measurements. In particular, we evidenced that magnetic easy axes flipped from out-of-plane [001] to in-plane [100] axis, followed by the <110> axes, with increasing hole density and temperature. Our study concluded on a general agreement with mean-field theory predictions of the expected easy axis reversals, and of the weight of uniaxial and cubic anisotropies in this material.

PACS: 75.50.Pp, 76.50.1g, 76.30.2v, 75.70.Ak

# INTRODUCTION

Long hailed as a possible room-temperature diluted semiconductor (DMS), $Ga_{1-x}Mn_xAs$ has yet to gain in Curie temperature, with a published record of 172 K[1]. It however proved to be a model material in the family of ferromagnetic DMS, yielding electrically[2] controllable magnetic properties, or low switching current densities in tunnelling magneto-resistance junctions [3,4].

In this material, the ferromagnetic phase arises from the exchange interaction between the localized manganese spins, and the delocalized carriers (holes) brought by the magnetic impurities. Its magnetic properties were historically described within the mean-field approximation, using a Zener-like model for the exchange integral[5]. Alternative theoretical models taking thermal and quantum fluctuations into account have also been proposed, and give distinct trends for the Curie temperature, or the saturation magnetization[6,7,8].

Experimental data have moreover evidenced a complex magnetic anisotropy[9,10]. Since demagnetizing effects are small in $Ga_{1-x}Mn_xAs$, the magnetic anisotropy is principally of magnetocrystalline origin, and directly reflects the anisotropy of the valence band (VB) through the spin-orbit coupling between magnetic impurities and carrier spins. As a result, magnetic easy axes vary with parameters controlling the shape and filling of the VB, such as the Zeeman splitting (via the temperature and the manganese concentration), the epitaxial strain, or the carrier density[11,12].

In this work, we have studied systematically the dependence of magnetic properties with carrier density, using magneto-transport experiments, magnetometry and Ferromagnetic Resonance (FMR). Our main objective was to compare our results to the predictions of the mean-field theory[11,12]. We investigated the Curie temperature, and the magnetic anisotropy of samples from low $10^{19}$ cm$^{-3}$ hole densities up to the highly metallic regime (low $10^{21}$ cm$^{-3}$), at fixed manganese concentration. Similar studies have already been reported, where authors used codoping[13,14], atomic-layer-epitaxy[15], post-growth annealing[16], or modulation-doped heterostructures[17] as means to decouple magnetic impurity and carrier concentrations.

We used another approach based on the hydrogen passivation of $Ga_{1-x}Mn_xAs$ layers. The diffusion of atomic hydrogen in the layer results in the formation of electrically inactive Mn-H complexes. This passivation reduces the carrier density by three orders of magnitude, and suppresses the ferromagnetic phase[18][19]. Resistivity measurements show that Mn-H complexes are stable for temperatures up to 100°C. It is still a matter of debate whether the hydrogen atom lies at a bond-centered or at an anti-bonding site between the acceptor and an arsenic neighbor[20][21][22]. According to density functional calculations[20], both of these configurations are most stable with the manganese in an S=5/2 state, the other spin configurations lying much higher in energy. This corroborates experimental findings that the manganese maintains 5 $\mu_B$ per atom after hydrogenation, in doped[19] ($x_{Mn} = 5.10^{20}$ cm$^{-3}$), or very diluted ($x_{Mn} = 10^{18}$ cm$^{-3}$) samples[23].

Upon controlled subsequent annealing, the hydrogen atoms leave the layer, restoring carriers to the matrix. By adjusting the annealing time, it is then possible to obtain samples with a constant manganese concentration, and hole densities ranging from $10^{17}$ cm$^{-3}$ to $10^{21}$ cm$^{-3}$.

Using this technique, we reported in a previous paper[25] the increase of the Curie temperature, and the easy-axis flip from out-of-plane to in-plane with increasing hole density, in qualitative agreement with mean-field predictions[12]. In this paper, we extended the study by performing magnetometry experiments and investigating in detail the magnetic anisotropy of our samples. To that end, we extracted the phenomenological anisotropy fields from the angular dependence of resonance fields in FMR experiments, at varying hole concentrations and for temperatures up to $T_c$. These anisotropy fields qualitatively explained the hysteresis cycles observed for magnetic fields applied both in-plane and out-of-plane at T = 5K. We also evidenced several easy-axis reversals with temperature and hole densities, in agreement with mean-field predictions.

# SAMPLE

A 50 nm ferromagnetic layer of $Ga_{1-x}Mn_xAs$ was grown on a semi-insulating GaAs (100) substrate. Manganese concentration was estimated by XRD to be around $x_{Mn}$ ~ 7 %[24]. An optimal annealing temperature was determined to both maximize Curie temperature, and stabilize mobile interstitial manganese atoms against future annealing: 1h at 250°C under a low $N_2$ flux yielded a Curie temperature of about 140 K. The layer was then passivated by a 130°C hydrogen plasma, cleaved, and annealed according to the procedure described in details in Ref. 25. Hall bars were also processed in order to perform magneto-transport experiments on each sample. The following study therefore focuses on a fully passivated sample, four partly depassivated samples with increasing hole densities (samples 1-4), and finally the reference sample that underwent the same exact thermal treatment as sample 4, but was hidden from the hydrogen plasma. Note that sample 4 was annealed long enough so as to contain almost no hydrogen. X-Ray diffraction measurements were performed on all samples (Table 1). As expected with a GaAs substrate, the layer showed compressive strain, with a bulk lattice mismatch varying from 0.36 % for the passivated sample to 0.17 % for the most annealed samples (4 and reference). The impact of these strain variations within the series will be discussed in the last part of this paper.

# MAGNETO-TRANSPORT AND MAGNETOMETRY

We first used magneto-transport Hall effect experiments to estimate the hole density $p$ of our layers, as a function of magnetic field H, sheet resistivity $\rho_{xx}$ and electron charge $e$ with $\rho_{xy} = H/pe + C\rho_{xx}^n M_\perp$ where $n=1$ or 2, $M_\perp$ is the perpendicular component of the magnetization, and C is a proportionality constant. Although a crucial parameter in most $Ga_{1-x}Mn_xAs$ studies, the precise determination of the carrier concentration is complicated by the dominating contribution of the second, anomalous Hall effect (AHE), term. It can however be estimated by saturating the magnetization at high magnetic fields and low temperatures[26]. Hole densities obtained by magneto-transport measurements in fields up to 10 T are given in Table 1. Values differed very weakly with temperature, when working at T=1.8 K, 4.2 K or 9 K, but were sensitive to the choice made for the sheet resistivity exponent, *n*, particularly at low carrier concentrations, where the magnetoresistance rose above 400 %. These data correlate fairly well to the mean-field predictions which yield a Curie temperature proportionnal to $p^{1/3}$ at fixed manganese concentration, and without taking into account the warping of the bands, as shown in Fig. 1.

|  | $T_c$ (K) | $p$ (cm$^{-3}$) | $\Delta a/a$ (%) | $H_c$ (Oe) |
|---|---|---|---|---|
| **Sample 1** | 42 [42] | 4.0 (2.0) x 10$^{19}$ | 0.31 | 140 |
| **Sample 2** | 70 [72] | 7.5 (3.2) x 10$^{19}$ | 0.30 | 66 |
| **Sample 3** | 83 [86] | 2.0 (1.1) x 10$^{20}$ | 0.27 | 50 |
| **Sample 4** | 130* [137] | 8.8 (7.9) x 10$^{20}$ | 0.18 | 20 |
| **Reference** | 140 [142] | 1.2 (1.0) x 10$^{21}$ | 0.17 | 12 |

**Table 1: Curie temperature determined by SQUID, magneto-transport (*) (Ref. 25), and FMR (measured at high magnetic field, between brackets); carrier density evaluated by Hall effect at high fields (4 T<H<10 T), and low temperature (T = 1.8 K) with n=1 (fit results for n=2 between parenthesis); bulk lattice mismatch deduced from XRD measurements, assuming GaAs Poisson coefficient ν=0.31; coercive field (± 15 Oe error) determined by SQUID at T = 5 K with H//<110>.**

We then performed magnetometry experiments using a QDMS superconducting quantum interference device (SQUID) magnetometer with the magnetic field lying in-plane along <110>

directions. For samples 1-3, and reference, we measured the temperature dependence of the sample magnetization under a 500 Oe applied field, after zero-field cooling (Fig 2). Comparison with field-cooled measurements showed no notable difference, thus excluding the presence of a second, super-paramagnetic phase in the sample. Also plotted for comparison is the Brillouin M(T) curve under a similar applied field, using $S_{Mn}$ = 5/2 and $T_c$ = 140 K. Curve shapes evolved notably with decreasing hole densities, becoming less and less convex. This feature was foreseen by mean-field approaches[12][27]. Indeed, when the hole density is low enough for the carriers to be entirely polarized before reaching the saturation magnetization, the molecular field seen by the manganese is not proportional to the magnetization anymore, and the M(T) curve ceases to follow a Brillouin function.

Using SQUID magnetometry with an in-plane applied field along <110>, we then measured the magnetization of our samples up to 5 T, at 5 K (Fig. 3). Coercive fields greatly decreased with increasing hole density (Table 1), going from $H_c$ = 140 Oe (sample 1) to 12 Oe (reference sample). Note that Potashnik et al[28] had already seen a decrease of $H_c$ with increasing $x_{Mn}$ but had not determined whether the critical parameter was the manganese or the hole concentration. We observed that magnetization at remanence was always smaller than at saturation, showing that, at 5 K, <110> is a hard axis at all hole densities. Strikingly, the least metallic sample (sample 1, $p \sim 3.10^{19}$ cm$^{-3}$) shows open hysteresis cycles for both the *in-plane* and *out-of-plane* field configurations (Fig. 3 & 4), with the easier axis along [001], evidencing a complex anisotropy for low hole/high manganese concentration samples, as we shall see later. After switching at $H_c$, the in-plane magnetization rotates in several steps for low $p$ samples (sample 1 and 2), and reaches saturation at high magnetic fields (~ 1 T, not shown on the figure). For the most metallic samples on the contrary (samples 3,4, and reference), it rotates much more continuously, and rapidly saturates at equivalent magnetizations (around 40 kA/m) for fields below 0.1 T. Taking 5 $\mu_B$ per Mn atom, this corresponds to $x_{Mn}^{eff}$ ~ 4.5%. The magnetization deficit compared to the nominal

Mn content is likely due to a small number of remaining interstitial or substitutionnal manganese frozen in an anti-ferromagnetic configuration.

Magneto-transport and magnetometry measurements clearly showed a complex magnetic behavior through the evolution of Curie temperature, M(T) curve shapes, and hysteresis cycles with hole density. In order to have a finer understanding of these phenomena, we then studied quantitatively the magnetic anisotropy of our samples using ferromagnetic resonance.

## FERROMAGNETIC RESONANCE

The ferromagnetic resonance (FMR) measurements were performed with an X-band spectrometer with standard 100 kHz field modulation and first derivative detection. All samples were measured in two different configurations, with ($\theta_H,\varphi_H$) the angles of the applied magnetic field H, and ($\theta,\varphi$) the equilibrium angles of the magnetization M. In general, H and M are no longer collinear with the exception of four high symmetry orientations, such that the equilibrium angles of M have to be calculated separately. In configuration 1 (in-plane) the magnetic field is applied parallel to the film plane, with $\theta_H = 90°$, and the variation of the FMR spectrum with the azimuthal angle $\varphi_H$ is measured. In configuration 2 (out-of-plane), $\varphi_H = \pi/4$ and H is varied from $\theta_H = 0°$, i.e. H//[001] to $\theta_H = 90°$, i.e. H// [110] or [1$\bar{1}$0]. As the absolute directions of [110] and [1$\bar{1}$0] have been lost during the fabrication process, we have assumed in accordance with previous results[29] that [110] is the harder axis of the two at 4 K. The FMR spectra are analyzed within the Smit-Beljers approach[30] based on the minimization of the free energy density F:

$$F = -MH \cdot [\cos\theta \cdot \cos\theta_H + \sin\theta \cdot \sin\theta_H \cdot \cos(\varphi - \varphi_H)] \quad -2\pi M^2 \sin^2\theta - K_{2\perp} \cos^2\theta$$

$$-\frac{1}{2} K_{4\perp} \cos^4\theta - \frac{1}{2} K_{4\parallel} \frac{(3+\cos 4\varphi)}{4} \sin^4\theta \quad -K_{2\parallel} \sin^2\theta \sin^2\left(\varphi - \frac{\pi}{4}\right)$$

(1)

The cubic magnetocrystalline anisotropy constant related to the zinc-blende structure of GaAs is $K_4$. The biaxial strain due to the lattice mismatch breaks the cubic symmetry, resulting in three phenomenological anisotropy constants: $K_{4//}$, $K_{4\perp}$, and $K_{2\perp}$. An additional magnetic anisotropy between [110] and [1$\bar{1}$0] axes is characterized by $K_{2//}$. The subscripts indicate parallel and normal to the film plane geometry respectively.

The first term of Eq. 1 represents the Zeeman energy, the second the demagnetization energy related to the shape anisotropy of the film, and the last terms the magnetocrystalline anisotropy. The FMR resonance condition for an arbitrary field orientation can be obtained from the Smit-Beljers equation[36]:

$$\left(\frac{\omega}{\gamma}\right)^2 = \frac{1}{M_S^2 \sin^2\theta}\left[\frac{\partial^2 F}{\partial\theta^2}\frac{\partial^2 F}{\partial\varphi^2} - \left(\frac{\partial^2 F}{\partial\theta\partial\varphi}\right)^2\right] \qquad (2)$$

where $\omega$ is the angular frequency of the microwave field and $\gamma$ the gyromagnetic ratio. The Landé g-factor of $Mn^{2+}$ is taken as g=2.00, independent of the hole concentration.

Typical resonance fields of these films vary between 1 kOe and 8 kOe. The FMR resonance fields measured for these two angular variations enable us to determine the numerical values of the four magnetocrystalline anisotropy constants from the resonance field positions of the high symmetry orientations H//[110],[1$\bar{1}$0], [001] and [100]. Considering a 10 Oe resolution on the position of the resonance fields, the precision of the anisotropy constants is mainly limited by the determination of the saturation magnetization ($\pm$ 10 %). These anisotropy constants vary strongly and differently with temperature in the 4 K to $T_c$ range. The FMR resonance position will also vary with the magnetization of the film, which depends on the hole concentration and thus on the hydrogen passivation. The numerical values of the magnetization M(T,H), which cannot be directly determined from the FMR measurements, have been obtained independently by the SQUID measurements for each sample.

The fully passivated sample does not show any FMR spectrum. It is no longer ferromagnetic but presents the typical exchanged narrowed EPR spectrum of a paramagnetic sample. The resonance field of 3300 Oe corresponds to a g-factor of g=2.04 (Fig. 5), signature of an $Mn^{2+}$ configuration for the manganese atoms. The hydrogenation process has therefore not modified the spin ground-state of the magnetic impurities. The competing interactions of dipolar line broadening and exchange narrowing have transformed the hyperfine split multiline spectrum of isolated $Mn^{2+}$ ions in a single, structureless Lorentzian line. It will not be further discussed here.

The annealed samples 1 to 4 are ferromagnetic. Fig. 6 shows typical FMR spectra at T = 20 K for the reference sample and the four partially passivated films (1,2,3, and 4) for the magnetic field orientation H//[001]. In addition to the dominant uniform mode at ~ 8 kOe, the reference sample shows some low intensity lines which we attribute to a sample inhomogeneity. We further observe a shift of the resonance fields to lower values with decreasing hole concentrations, which reflects the change of the anisotropy constants with *p*. Note that the linewidth increases in the same manner. The linewidth is related to the damping factor as well as to sample inhomogeneities.

From the resonance fields for H//[001], [100], [110], [1$\bar{1}$0] and the magnetization value M, we have determined the four anisotropy constants (Table 2) and simulated the complete angular variation of the resonance fields. Fig. 7 shows these angular variations for all samples in both configurations, for a fixed temperature T = 4 K. The magnetic easy axes directions correspond to the lowest resonance fields for each sample.

At 4 K, the easy axis is in-plane along <100>, except at very low carrier concentration (sample 1), where it lies out-of-plane. Indeed, FMR spectra and magneto-transport experiments evidenced an easy axis reversal from out-of-plane to in-plane between $p \sim 3.10^{19}$ cm$^{-3}$ (sample 1) and $p \sim 5.10^{19}$ cm$^{-3}$ (sample 2). This is in part due to the fact that below a particular Fermi energy, carriers lie mainly in the heavy-hole sub-band, favoring the alignment of the spins along the growth direction[9]. Mean-field calculations using an effective $x_{Mn}$ = 5% and $\varepsilon_{xx}$ = - 0.2 % estimated a critical hole density of $p_c = 7.10^{19}$ cm$^{-3}$ [31], close to our experimental values, and those found by Sawicki. et. al[9]. We further observed for all samples a nonequivalence of the [1$\bar{1}$0] and [110] directions.

|  | $H_{2\perp}$ (Oe) | | $H_{2//}$ (Oe) | | $H_{4\perp}$ (Oe) | | $H_{4//}$ (Oe) | |
| --- | --- | --- | --- | --- | --- | --- | --- | --- |
|  | T=4K | T/T$_c$=0.7 | T=4K | T/T$_c$=0.8 | T=4K | T/T$_c$=0.8 | T=4K | T/T$_c$=0.8 |
| **Sample 1** | 2373 | -193 | 164 | 134 | 341 | -256 | 1815 | 444 |
| **Sample 2** | -1551 | -1375 | 208 | 108 | -137 | -19 | 1745 | 27 |
| **Sample 3** | -2259 | -976 | 163 | 31 | -678 | -231 | 1099 | 11 |
| **Sample 4** | -3311 | -854 | 200 | -70 | -908 | -511 | 342 | -5 |
| **Reference** | -3178 | -1099 | 261 | -30 | -1283 | -97 | 362 | -4 |

**Table 2: Anisotropy fields extracted from FMR measurements for samples of increasing carrier concentration. First column: T = 4K, second column: T/T$_c$, with T$_c$ determined by FMR (see Table 1).**

When the temperature increases, the perpendicular easy axis of sample 1 flips to [100] at T = 10 K, as has also been observed previously[9][10]. This was verified by the closing of the hysteresis cycles in Hall effect measurements (not shown here), and can be explained by a band-filling argument similar to that used for the easy axis reversal with *p* (see above). While keeping in mind that absolute axes orientations were determined by supposing [110] harder than [1$\bar{1}$0] at 4 K, we observed that for samples with intermediate hole concentrations (samples 2 and 3), the in-plane [100] easy axis switched to [1$\bar{1}$0] for $T/T_c$ = 0.8. In the most doped samples on the contrary (sample 4 and reference), the in-plane [100] easy axis switches to [110] for $T/T_c$= 0.8. While temperature-induced reorientations from [100] to <110> had already been observed in thicker and/or less doped samples[32][16], we show here that this easy-axis reversal spans a whole order of magnitude in hole concentration, at fixed manganese density.

## DISCUSSION

**Anisotropy fields**

Anisotropy fields found by FMR at T = 4 K (Table 1) scale reasonably well with experiments done on comparable $Ga_{1-x}Mn_xAs$ layers[14][33][34]. The anisotropy fields $H_i$ are calculated from the anisotropy coefficients $K_i$ via the magnetization: $H_i=2K_i/M$. Note that some authors refer to $K_{cubic}$ as $K_4=K_{4//}= K_{4\perp}$, and to $K_{uniaxial}$ as $\pm K_{2\perp}$.

In our series, the general trend is a monotonous evolution of anisotropy fields with carrier density $p$. The most remarkable variations are seen for the cubic terms: $H_{4//}$ is divided by six, while $H_{4\perp}$ is multiplied by four when $p$ is increased by an order of magnitude. Contrary to another study[17], the approximation $|H_{4\perp}|<<|H_{4//}|$ is far from valid in our samples. $H_{4//}$ and $H_{4\perp}$ are indeed systematically different by a factor of three at least, and change in relative signs and values with increasing hole densities.

At 4 K, the planar anisotropy is at all carrier densities dominated by the cubic term $H_{4//}$, and decreases with hole density. It is a result of the in-plane anisotropy of the valence band, as was shown by Abolfath et al.[11] and Dietl et. al[12]. Calculations implementing a mean-field approach that does not take into account the in-plane uniaxial contribution ($H_{2//}$ here), showed that the in-plane cubic anisotropy increases with decreasing hole density, up to a critical carrier concentration ($2.10^{19}$ cm$^{-3}$ considering x$_{Mn}$=5%, T=0 K, and J$_{pd}$ = 50 meV.nm$^3$) where it then starts to decrease[11]. The high *p* predictions scale qualitatively well with our data, but our samples were too doped to investigate the very low density regime. We observe that the in-plane cubic anisotropy (H$_{4//}$) dominates at low temperature, and the uniaxial anisotropy (H$_{2//}$) at high temperature, as also suggested in Ref. 9.

The largest anisotropy term is always clearly the perpendicular uniaxial H$_{2\perp}$ field, which can be up to ten times higher than the in-plane uniaxial anisotropy, H$_{2//}$. H$_{2\perp}$ increases with *p,* as also observed elsewhere (Refs. 35 36), whereas H$_{2//}$ stays at a constant ~ 200 Oe along the series (at 4 K). Uniaxial anisotropies decreased with temperature, after reaching a peak at around 20-30K. The perpendicular H$_{2\perp}$ term was the only one to remain large up to T$_c$, converging to a value of about 1000 Oe for all samples. These observations seem to confirm that Ga$_{1-x}$Mn$_x$As can, at the lowest order at least, be considered a uniaxial ferromagnet.

Finally, note that anisotropy fields for samples 4 and reference were similar within 30%. Moreover, magneto-transport experiments had shown in Ref. 25 that these two samples had similar Curie temperatures and magnetization curves M(H). This tends to confirm that our hydrogenation is indeed a non-destructive and reversible process, since we retrieve structural and magnetic parameters close to the initial ones after emptying the layer of its hydrogen by thermal annealing.

**Magnetization reversal process**

Following the work of Liu et. al.[37], we then calculated the free energy density per Mn atom as a function of the magnetization polar angles $\Theta$ and $\phi$, assuming no thermal fluctuations, and anisotropy coefficients K$_i$ independent of the applied magnetic field. In order to compare correctly the free energy to the thermal energy k$_B$T, it would in fact be more appropriate to compute either the free energy of all manganese atoms (considering a coherent magnetization reversal), or of the nucleation volume (if considering a nucleation/propagation reversal mechanism). At this point of the study however, we cannot give a reasonable value for this critical volume, and therefore prefer to plot the free energy density per Mn atom, keeping in mind the former remark. We used Equation (1) and the magnetization given by the M(T) SQUID curves, taking 5 $\mu_B$ per atom. Note that we took into account the contribution of the demagnetizing field, but that it represented less than 15% of the total energy and had little impact on the conclusions.

**Magnetic field lying in-plane**

For sample 1 with H//<110> (Fig. 8.a), two equivalent valleys arise with decreasing magnetic fields, on either side of $\phi = 45°$. Note that for $H_{applied} = 0$ Oe, [110] and [1$\bar{1}$0] are not equivalent, an indication of the uniaxial in-plane anisotropy ($H_{2//}$). The angles minimizing the energy indicate that at H = 2200 Oe, the magnetization starts to turn slowly away from [110] to the easier [100] axis, flips abruptly around H ~ 250 Oe to [0$\bar{1}$0], and rotates slowly again towards [$\bar{1}\bar{1}$0]. This is qualitatively what is observed by SQUID magnetometry, with a multi-step magnetization-reversal (Fig. 3).

When the carrier density increases, the anisotropy fields evolve, and give a quite different magnetization reversal process. For sample 4 (Fig. 8.b) for example, we see that the magnetization rotates very progressively. While in low *p* samples, the large positive $H_{4\bar{i}//}$ fields (>1000 Oe) strongly favor <100> axes over the magnetic field direction <110> and yield high coercive fields, the low $H_{\bar{4}//}$ values in high *p* samples are responsible for the smooth magnetization rotation and low coercive fields. Indeed, when the hole density increases, the planar anisotropy progressively diminishes, until in-plane <110> and <100> axes become almost equivalent.

**Magnetic field lying out-of-plane**

For sample 1 ($p \sim 3.10^{19}$ cm$^{-3}$), computing the free energy with H//[001] (Fig. 9.a) shows that the magnetization first remains collinear to the direction of the decreasing magnetic field, then flips abruptly in the opposite direction around 400 Oe: [001] is a magnetic easy axis. This is indeed what is observed in Hall effect hysteresis cycles (Fig. 4), with $H_c^{exp} \sim 300$ Oe. A similar argument applied to sample 2 ($p \sim 5.10^{19}$ cm$^{-3}$) can explain its unusual experimental hysteresis cycle. Energy curves (Fig. 8.b) show that in this case, the magnetization first stays collinear to the [001] direction, then abruptly drops to the easier [101] axis ($\theta = 45°$) at H = 1950 Oe, before rotating continuously towards the next axis [10$\bar{1}$] ($\theta = -45°$), and finally flipping completely perpendicular-to-plane at high magnetic field. The difference in energy between [001] and [101] configurations gives a characteristic jump at $H^{exp} = 1800$ Oe in the hysteresis cycle (Fig. 3).

On the remaining samples, we can show in the same way that anisotropy fields obtained by FMR render fairly well hysteresis cycles for all field configurations, as has already been observed in samples with lower manganese doping[37]. Agreement with experimental reversal fields is surprisingly good given the strong hypotheses of this model. We therefore concluded on the validity of this approach to study the magnetic anisotropy of our samples.

After having investigated a layer grown in compressive strain, we started a second study on a 50 nm Ga$_{0.93}$Mn$_{0.07}$As layer grown in *tensile strain*, over a Ga$_{1-y}$In$_y$As buffer. In this case, the biaxial strain yields a perpendicular-to-plane easy axis for the reference, highly doped layer. Growth details are given in Ref. 38. Using the same procedure as in Ref. 25, a set of samples with hole concentrations ranging from $10^{18}$ cm$^{-3}$ to about $10^{21}$ cm$^{-3}$ was fabricated, with identical manganese content, and Curie temperatures of 35 K, 45 K, 70 K, 95 K and 156 K (determined by transport measurements). We observed at low temperature an in-plane to out-of-plane easy axis transition with increasing hole density. Moreover, all low *p* samples recovered a perpendicular-to-plane easy axis with increasing temperature, with a transition around 20 K. These observations fit very nicely to the qualitative conclusions of mean-field calculations for layers in tensile strain (see Fig. 10 of Ref. 9 for example). Although this study remains preliminary, it has proved that our hydrogenation method is valid for layers in both compressive and tensile strain, and has yielded promising results concerning the evolution of magnetic anisotropy with hole density for (Ga,Mn)As layers in tensile strain.

## CONCLUDING REMARKS

In the context of competing theories concerning DMS, and in particular, Ga$_{1-x}$Mn$_x$As, experimental results concerning the evolution of Curie temperatures and magnetic anisotropies with carrier density are an important test. We developed an original technique using hydrogen passivation to tune the carrier density only, keeping the structural parameters reasonably constant. It can be argued that the strain discrepancies in our series of samples (Table 1) may have influenced the magnetic anisotropy, as suggested by Dietl et al.[12]. If this were valid, we would however expect a *decrease* of the perpendicular anisotropy terms in the series while we observed experimentally the opposite trend (see Fig. 9 of Ref. 12). We therefore assumed the strain differences to have little impact on our study.

Using the anisotropy fields deduced from the FMR measurements we showed that the shapes of the hysteresis cycles were due to a complex combination of cubic and uniaxial, perpendicular and planar terms. We found that cubic terms $H_{4//}$ and $H_{4\perp}$ were quite different, and that the uniaxial in-plane term $H_{2//}$ was far from negligible, being of the order of 200 Oe at 4 K, and varying very little with $p$. We corroborated mean-field predictions concerning the increase of the uniaxial perpendicular term $H_{2\perp}$, and the decrease of the in-plane anisotropy field $H_{4//}$ with hole density. We confirmed mean-field predictions of easy axis reversals from [001] to [100], to <110> axes, with increasing carrier density, and/or temperature. Lastly, the Curie temperature was indeed found to be proportionnal to $p^{1/3}$ in good approximation, over two orders of magnitude of hole densities. We therefore conclude on the general experimental agreement with mean-field predictions for $Ga_{1-x}Mn_xAs$ in compressive strain.

## ACKNOWLEDGMENTS

This work has been supported by the Région Ile de France, the Conseil Général de l'Essone, and through the ACN Jeunes Chercheurs BOITQUAN and the PNANO MOMES. We thank R. Mattana, R-M. Galéra and Ph. Monot for access to their SQUID facility. Finally, we thank J. Zemen for his calculations, and insightful remarks on our study.

**FIGURES:**

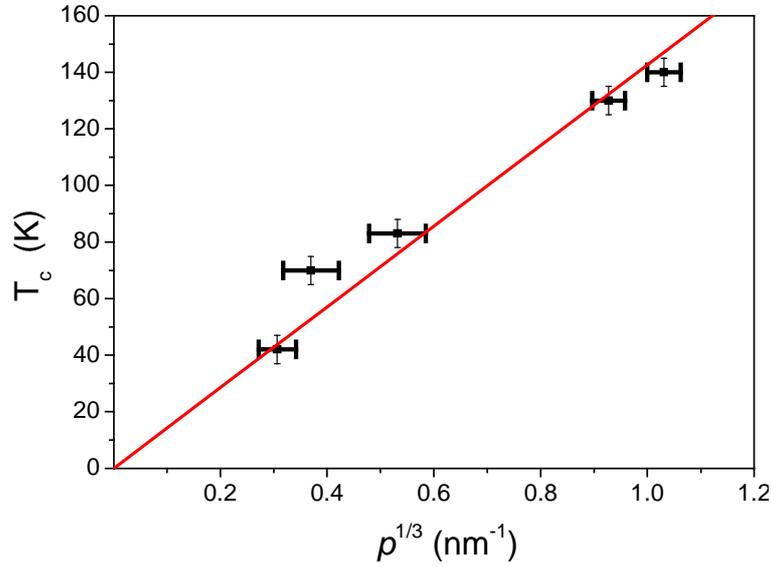

**Fig. 1: (Color online) Curie temperature as a function of $p^{1/3}$ where $p$ is the mean hole density between $n=1$ and $n=2$ fit results (given by the tips of the horizontal bars). Full line is the fit to the mean-field expression $T_c \propto p^{1/3}$.**

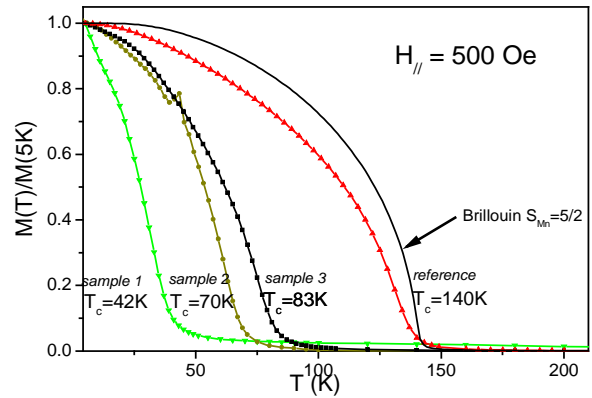

**Fig. 2: (Color online) Temperature-dependence of the normalized magnetization under a 500 Oe in-plane field for samples 1-3, and reference. The solid line corresponds to Brillouin curve with $S_{Mn} = 5/2$, and $T_c = 140$ K. Curie temperatures increase and curves become more convex with increasing hole density.**

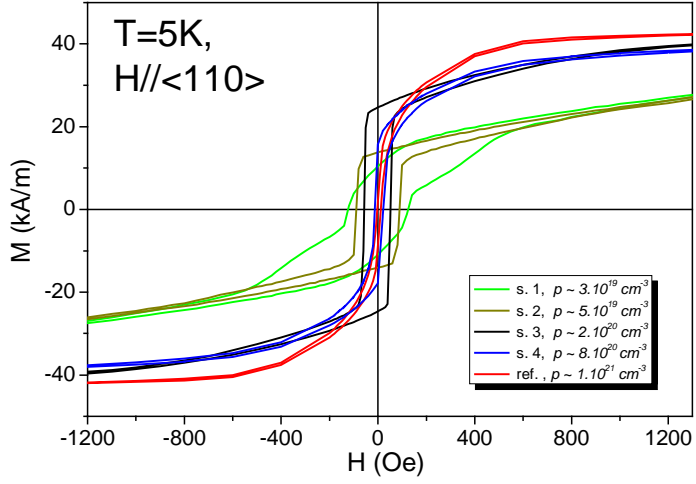

**Fig. 3: (Color online) Hysteresis cycles obtained by SQUID magnetometry with the magnetic field lying in plane.**

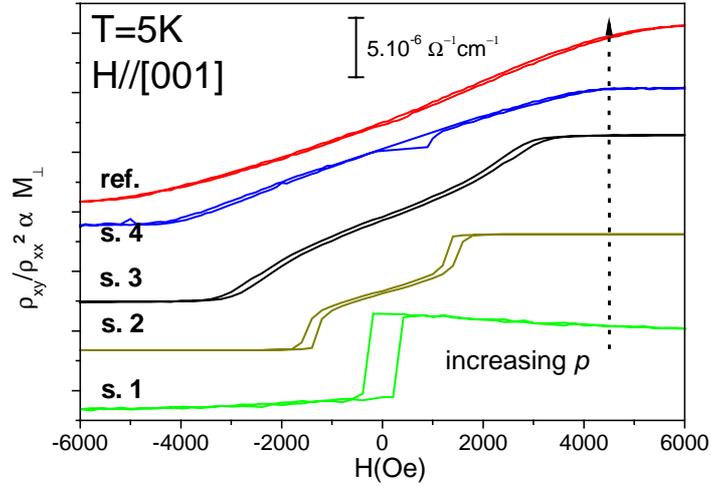

**Fig. 4: (Color online) Hysteresis cycles obtained by anomalous Hall effect[25] with H//[001] (off-set for clarity), and assuming $M_\perp \propto \rho_{xy}/\rho_{xx}^n$ with $n=2$. The magnetic easy axis flips from out-of-plane to in-plane with increasing hole density.**

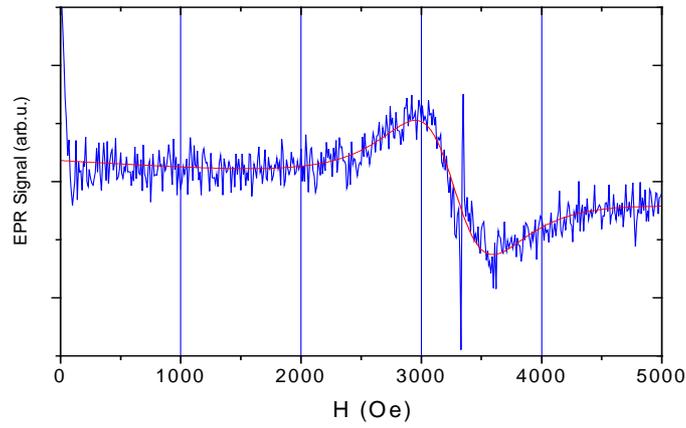

**Fig. 5: EPR spectrum of the fully passivated sample; T=20K. The sharp line at 3300 Oe is a surface signal.**

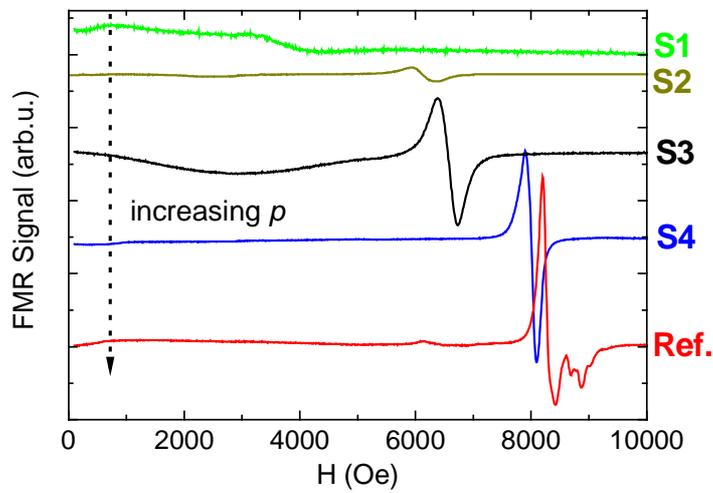

**Fig. 6: (Color online) FMR spectra at T = 20 K and H//[001] for the reference sample and samples 1-4; the spectra are measured with the same gain.**

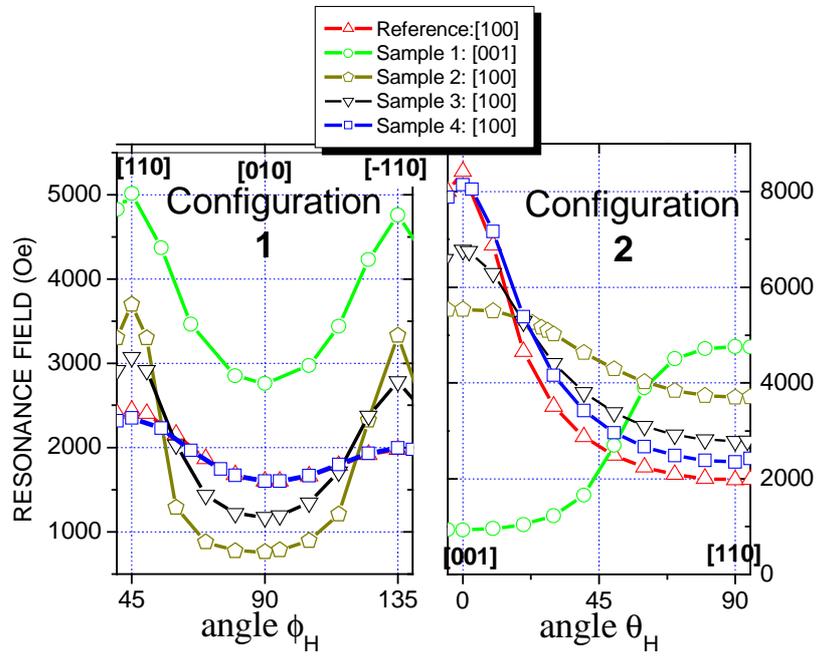

**Fig. 7:** (Color online) Angular variation of the FMR resonance fields at 4 K for (1) "in-plane" configuration, and (2) "out-of-plane" configuration for samples 1-4 and reference. Magnetic easy axes are indicated in the caption. Symbols: experimental results, lines: simulated variation with the coefficients given in Table 2.

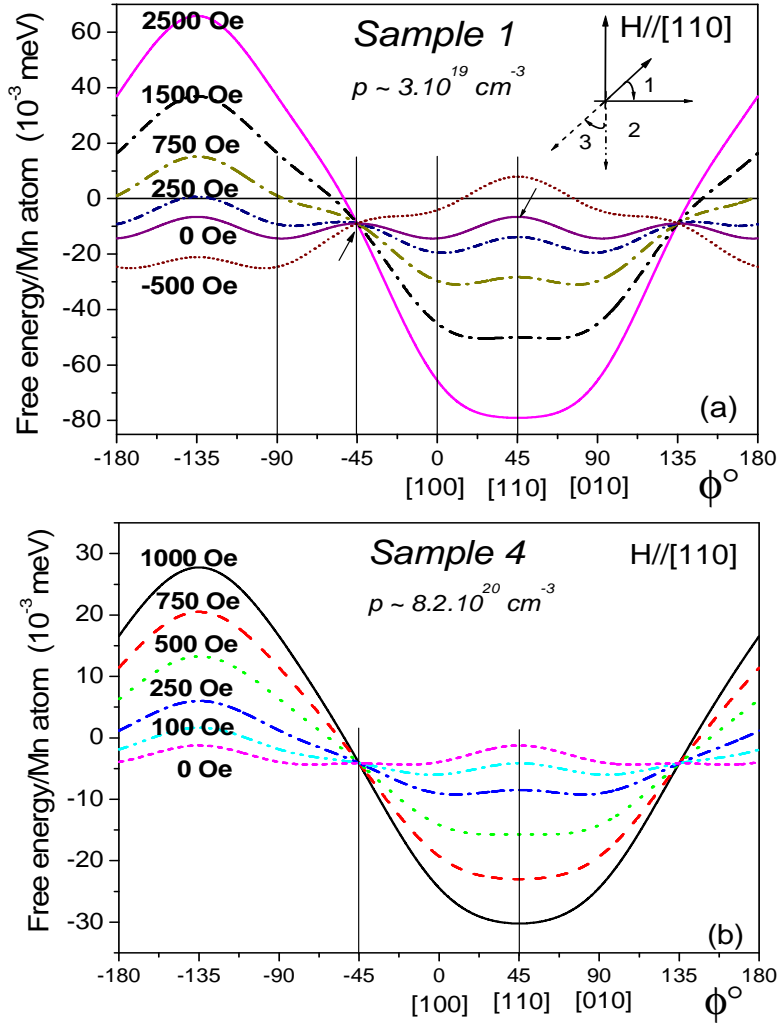

**Fig. 8: (Color online) Free energy per Mn atom computed with anisotropy fields given by FMR experiments at T=4 K; the magnetic field is applied in-plane along [110]. For sample 1 (a), the magnetization rotates in 3 steps, showing a competition between <100> and <110> axes. The arrows indicate the two non-equivalent magnetization configurations along [110] and [1-10], resulting from the small in-plane uniaxial anisotropy $H_{2//}$. For sample 4 (b), the magnetization rotates continuously, <110> is a hard axis.**

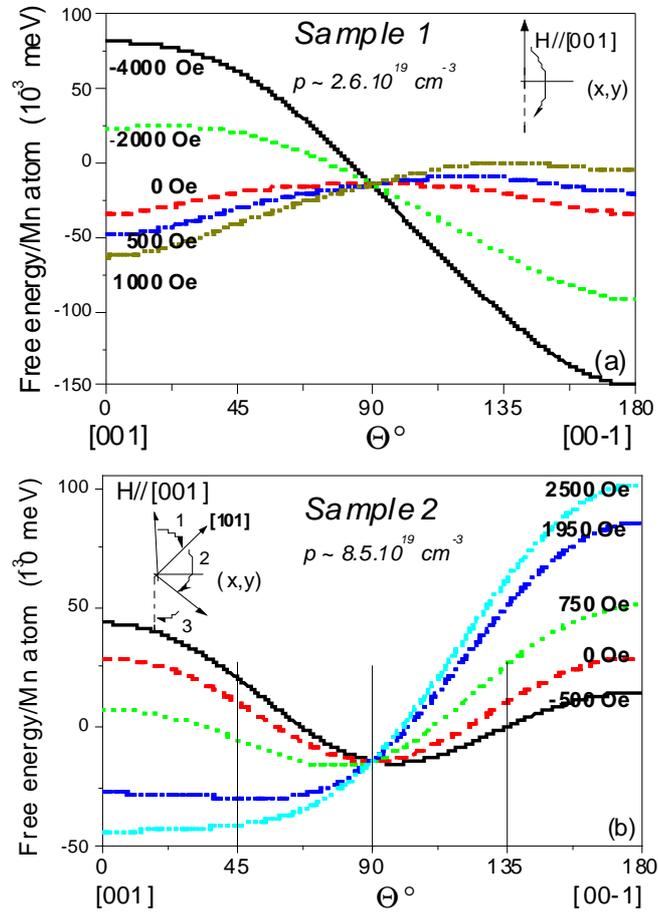

**Fig. 9: (Color online) Free energy per Mn atom computed with anisotropy fields given by FMR experiments at T=4 K; the magnetic field is applied perpendicular-to-plane. At low hole densities (a), the easy axis is [001]. Upon increasing  $p$  (b), an in-plane component of the magnetization appears.**